\documentclass[aps,floats,amssymb,twocolumn,prb]{revtex4-2}
\usepackage{calc}
\usepackage{psfrag}
\usepackage{graphicx}
\usepackage{color}
\usepackage{bm}
\usepackage{float}
\usepackage{lipsum}
\usepackage{amsmath}
\usepackage[unicode,linktocpage=true]{hyperref}
\hypersetup{colorlinks=true,linkcolor=blue,urlcolor =blue,citecolor=blue,anchorcolor=blue} 

\def\bs{\boldsymbol}

\begin{document}

\title{Pair density wave in the fractional quantum Hall effect at even denominator}

\author{M.V. Milovanovi\'c}
\affiliation{Scientific Computing Laboratory, Center for the Study of Complex Systems,Institute of Physics Belgrade, University of Belgrade, Pregrevica 118, 11080 Belgrade, Serbia}

\begin{abstract}
The fractional quantum Hall effect (FQHE)  at filling 5/2, which is usually understood as a $p$-wave paired state of underlying quasiparticles - composite fermions, transforms into a nematic phase under pressure \cite{csathy0, csathy}. A pair density wave (PDW) may be a precursor, underlying state for this behaviour, and such state(s) were proposed that maintain  the weak-pairing feature of the uniform paired state \cite{frad}. Based on considerations in the weak-coupling regime of a microscopic description of the pairing phase (to mimic the phase as it gives way to a nematic phase in the experiments), we argue that the ensuing and relevant PDW state has a strong-pairing character. Furthermore, due to the existence of a single collective mode associated with the order parameter in the uniform paired phase, in the weak-coupling regime, the $p$-wave paired state, in general (for example, in the superconducting state of electrons), may be prone to a PDW instability.
\end{abstract}

\maketitle

\section{Introduction}
Pfaffian states are well-known FQH states, believed that represent the essence of the physics at filling factor $5/2$; they are the model ground state wave functions, which are very close to the exact ones. The Pfaffian i.e. Moore-Read (MR) state as proposed in \cite{mr}, can be described as a collection of Cooper pairs of underlying quasiparticles - composite fermions (CFs). All electrons i.e. CFs are involved and the construction can be identified  as a description of a strong-coupling regime of a $p$-wave topological superconductor (SC). The MR state can be connected with the FQH physics and ground state for the Coulomb problem in the second Landau level (LL) at filling factor $5/2$; the overlap with the exact ground state in the numerical experiments is considerable \cite{rh}. If the $V_1$ Haldane pseudopotential is increased from its Coulomb value, that would increase the pay for the state of two electrons or CFs, and thus decrease the likelihood of pairing. In the numerical experiments \cite{rh} it was noticed  (in some geometries)  that without a closing of the gap of the paired state, there are crossings among excited states, as the $V_1$ pseudopotential is increased. Thus we may speak of a possibility where low-lying spectrum is rearranged though the nature of the topological state is not changed. This evolution must correspond to a crossover transition from the strong to weak coupling of the weak $p$-wave pairing  phase of CFs. The rearrangement of excited states should correspond to the disappearance  of non-Abelian quasiparticles  - half-flux (where unit of the flux is $ (h c)/ e $ ) vortices from the low-lying spectrum and entering into the phase that corresponds to the SC state of the I kind of CFs \cite{pm}. (Namely the evolution with increasing $V_1$ is followed by increasing likelihood  of Fermi liquid of CFs, and the physics of localized, collective excitations that can be considered as  a consequence of insertions of flux quanta, is that of ordinary Laughlin quasiparticles (quasiholes)). This phase in the lowest lying spectrum, in the bulk, supports only Laughlin quasiparticles or quasiholes with the characteristic charge equal to $1/2$ of the unit of electric charge.

In the following we will use phrase "weak-coupling phase of Pfaffian" to denote the weak-coupling $p$-wave  SC phase of CFs. If, in an anology with the physics of a SC of the I kind we assume a disappearance of vortices from the spectrum, and their existence only as a part of macroscopic domains, the described weak-coupling phase of Pfaffian would not exist as a real phase, but as a critical theory next to phase separated systems (with macroscopic regions with different densities) if the system is away from half-filling \cite{pm}. But we must consider also a second possibility, because our system is not a simple SC of (neutral - dipole) CFs, but also involves charge degrees of freedom, which may preclude the clumping of vortecies i.e. charged excitations in the bulk, except for pair of vortices in the low lying spectrum, which will fuse into the Laughlin quasiparticles. This weak-coupling phase of CFs may be stable towards phase separation.  In any case, either as a critical or real phase of weakly-coupled CFs, its description may serve as a starting point in understanding of the appearance  of the PDW phase as we will describe in the folowing.

Such a PDW phase maybe a precursor state for the nematic phase observed in experiments in the FQHE at filling factor $5/2$ \cite{csathy}. This is proposed in Ref. \cite{frad}, which thoroughly analyses a weak-pairing scenario for a PDW state. Here we will take a different, microscopic route, based on a BCS Hamiltonian that describes the weak-coupling regime of pairing and argue that a resulting PDW state characterizes strong-pairing of CFs. Thus we are proposing an alternative scenario for a PDW state, which assumes a weak-coupling regime of CFs, before entering a PDW phase with more usual, strong-pairing correlations.

In the following section, Section II, an effective descrition of the weak-coupling regime of the Pfaffian phase is proposed and justified. In Section III, an appropriate pairing operator is introduced, which Heisenberg equations of motion are analyzed in the next section, Section IV. In Section V, the description of the collective behaviour and mode in the previous section is used to explain the mechanism leading to a PDW state. Section VI is devoted to a discussion and conclusions.

\section{Effective description of weak coupling Pfaffian}
We would like to consider and analyze collective modes in the Pfaffian weak-coupling regime.  The collective modes in the strong coupling regime are thoroughly analyzed together with the phenomenon of supersymmetry of modes, associated with the low-lying Majorana and non-Abelian quasiparticles in \cite{ss}. Our ultimate goal is to describe a transition from the Pfaffian weak coupling phase into a pair density wave (PDW) of CFs, and to show that the physics of the Pfaffian weak coupling phase i.e. regime is amenable to the establishment of such a PDW. (This phase  with the PDW instability of CFs, and, as we will find out, in the strong pairing regime, is not the one previously discussed in the context of nematic physics at half-filling   \cite{frad}, which maintains weak pairing.)

Firstly, we would like to consider a model Hamiltonian that we find appropriate to describe the strong to weak coupling transition of the Pfaffian phase. The Hamiltonian for CFs has the following form
\begin{equation}
H ={\cal  H} + \int \frac{d{\bs{k}}}{(2\pi)^2} C \exp(- \frac{q^2}{2})  \rho^{R}({-\bs{q}})
  (\rho^{R}({\bs{q}}) - \rho^{L}({\bs{q}})), \label{modHam}
\end{equation}
where,
\begin{eqnarray}
{\cal H} =\frac{1}{2} \int \frac{d{\bs{q}}}{(2\pi)^2} \;{\tilde V} (|{\bs{q}}|)   \label{RHam}
: \rho^{L}({\bs{q}})
 \rho^{L}(-{\bs{q}}):.
\end{eqnarray}
In the context of the boson system at filling factor one, it can be justified in a microscopic fashion, by working with the constraint $ \rho^{R}({\bs{q}}) = 0$ for the "unphysical" degrees of freedom \cite{read}.  While 
\begin{equation}
\rho_{\bs{q}}^{L} = \int \frac{d\bs{k}}{(2\pi)^2} c_{\bs{k} - \bs{q}}^\dagger c_{\bs{k}} \exp\left(i \frac{\bs{k} \times \bs{q}}{2}\right), \label{BDen}
\end{equation}
represents the density of "physical" degrees of freedom, where $  c_{\bs{k}}^\dagger ,  c_{\bs{k}} $ are creation and annihilation operators for CFs,
\begin{equation}
\rho_{\bs{q}}^{R} = \int \frac{d\bs{k}}{(2\pi)^2} c_{\bs{k} - \bs{q}}^\dagger c_{\bs{k}} \exp\left(- i \frac{\bs{k} \times \bs{q}}{2}\right) \label{BDenr},
\end{equation}
represents the density of "unphysical" degrees of freedom - correlation holes - well developed (i.e. localized, charge quantized) entities in the FQHE. Thus we work in an enlarged space with an appropriate constraint.

The first part of the Hamiltonian for CFs in (\ref{modHam}) is the second-quantized form of the interaction operator inside a LL. The details of a derivation can be found in \cite{read}. The states of the system in the enlarged space, the physical states, are defined by the constraint $ \rho^{R}({\bs{q}}) = 0$, and the construction in (\ref{modHam})  reduces to the interaction operator, on the physical space.  (There is no kinetic term inside a LL.)  The extra term in  (\ref{modHam}), the null operator, is necessary (as a way of adapting the form of the Hamiltonian) to produce the $p$-wave instability, in the scope of the mean field approach. Based on the numerical experiments, it is known that the system of repulsive bosons at filling factor one is prone to pairing and a way to reproduce the pairing at the mean-field level is to use  (\ref{modHam})  \cite{nnmvm}. The form of this extra term is completely physically justified. It emphasizes the importance of the dipole representation, which exposes the fact that, together with electrons, correlation holes are built in  the qusiparticle - CF description. Then, the cause of the pairing is described by this extra term, which represents monopole-dipole interaction;  an electron is attracted  to another which is followed by a (correlation) hole. The resulting pairing is on the level of quasiparticles - CFs . The constant $C$ has to be greater than $2$ in order for $H$, in a mean-field treatment,  to describe $p$-wave pairing (of CFs), and large enough $(C \gtrsim 30) $ to describe the strong-coupling pairing regime  of CFs \cite{nnmvm}. We choose $C$  as a constant, to describe in a minimal, effective way the transition between the strong- and weak-coupling regime though in the numerical experiments \cite{b3}, microscopic interaction has a structure with $V_0 $ and $V_2$ Haldane pseudopotentials, and as $V_2$ gains importance, we are getting near the optimal - strong-coupling Pfaffian.

We would like to add that the extra term in  (\ref{modHam}) can be mapped via  the usual Chern-Simns transformation into the underlying bosonic system representation, and, in that mapping, the monopole-dipole interaction of CFs, becomes the basic 3-body (two delta-function) interaction among bosons \cite{nnmvm}, which is known to be a model interaction for the Pfaffian state in that system.

We will adopt the form of the Hamiltonian given in (\ref{modHam})  to describe the weak-coupling regime, and the transition from the weak to strong coupling regime, also for the electron system that occupies half of the available states in a LL (a half-filled  LL system). It will be a model, effective Hamiltonian,  for the system at filling factor $5/2$, in the second LL. Thus, as in the bosonic case, the microscopic interaction, in the strong coupling limit, may have a structure (a set of Haldane pseudopotentials), but we will model its influence (in the weak coupling regime) through parameter $C$.

In other words, to describe the weak coupling regime of Pfaffian pairing, which will be our main focus in the following, we will use (\ref{modHam}) with that extra term with constant C as a minimal model to induce and describe pairing (and the transition to the strong-coupling regime).  For a more detailed description of the strong-coupling regime we need to invoke a critical Hamiltonian and microscopic interaction to describe relevant parameters that would lead to Fermi liquid and paired states as in \cite{nnmvm}.

If the presence of the term with parameter $C$  can be easily introduced and justified (due to the constraint) in the bosonic system, the presence of the parameter $C$  in the electron system, if necessary, can be justified by allowing (static) screening in the real systems (which is absent in a fixed LL).

\section{Pairing operator}
 To study the collective modes in a BCS system, we may examine the Heisenberg equations of motion of the pairing operator,
\begin{equation}
\delta({\bs{q}}) = \int \frac{d\bs{k}}{(2\pi)^2} c_\uparrow ({-\bs{k} + \bs{q}}) c_\downarrow ({\bs{k}}) \label{po} ,
\end{equation}
which represents the Fourier transform of the operator we associate with the order parameter. In (\ref{po})  $c_\sigma$' s represent annihilation operators for fermions (electrons) that pair and $\sigma$ is for spin.  Here we discuss pairing of fermions without spin, and furthermore they live in a LL.  Also the fermions are composites and if we consider their microscopic description, it is done via fermionic operators with two indexes \cite{read,ph}, one index denoting the state of electron (boson),  and the other, the state of its correlation
hole. The fermionic bilinears can refer also to two indexes if we trace out other two indexes. In this way we can define operators which correspond to densities of electrons (''physical density") or correlation holes ("unphysical density"). But also we can consider bilinears with two annihilation operators or two creation operators of fermions and trace out indexes of mixed kind to get a pairing operator of the following form, 
\begin{equation}
\eta ({\bs{q}}) = \int \frac{d\bs{k}}{(2\pi)^2} i \sin \left( \frac{\bs{q} \times \bs{k}}{2}\right)  c_{-\bs{k} +\bs{q}} c_{\bs{k}}. \label{ppo}
\end{equation}
 We introduce
\begin{equation}
f(\bs{q}, \bs{k}) = i \sin \left( \frac{\bs{q} \times \bs{k}}{2}\right).
\end{equation}
The details of the derivation of the pairing operator $\eta ({\bs{q}}) $ can be found in the Appendix.
In  polarized systems, a simple generalization of the pairing operator in (\ref{po}) does not exist, but in the case of CFs, a generalization as in (\ref{ppo}) is intimately connected with the composite structure of the fermions and that any pairing instability is followed by a correlation - an excitonic attraction between electrons and correlation holes. Thus a summation over mixed indexes ("physical"  and  "unphysical"-correlation hole) leads to the pairing operator. The CF representation endows a status of physical degrees of freedom to correlation holes. See also a discussion in the Appendix. The $\eta ({\bs{q}}) $ operator is a generalization of $\delta({\bs{q}})$ as a permissible - non-trivial superposition of Cooper pairs with non-zero momentum of the center of mass, but also with special coefficients inside the sum - they reflect the projection to the space of a LL. 

Furthermore, each pair in $\eta ({\bs{q}})$ describes a composite - a quadrupolar excitation, because each CF represents a dipole. The exact structure of the dipole in the strong coupling limit is connected with the effective constraint that is valid in that limit \cite{nnmvm},
\begin{equation}
\rho^{L}({\bs{q}})  + \rho^{R}({\bs{q}})  = 0.
\end{equation}
It is not hard to prove (along the same lines as for the $\rho^{R}({\bs{q}}) =\ 0$ constraint in \cite{dose}) that opposite charges in the dipole are charges of $1/2$ unit of charge and thus "partons" or "semions", connected with basic Laughlin excitations in a half-filled LL. These pairs of semions are  at the heart  of the Dirac picture  of the half-filled LL \cite{csen}. On the other hand, away from this strong-coupling regime,  in the case of bosons, the  microscopic constraint is valid i.e. 
 $\rho^{R}({\bs{q}}) =\ 0$, and dipole consists of a boson and its correlation hole of unit, opposite charges. But in the case of the half-filled level of electrons, the constraint  $\rho^{L}({\bs{q}}) + \rho^{R}({\bs{q}}) =\ 0$ is in effect (not only in the critical region); it is based on the expectation of the particle-hole symmetry and other phenomenological reasons as detailed in \cite{pkm}. Thus, in the context of the usual systems of electrons at half-filling, the quadrupolar excitations associated  with  $\eta ({\bs{q}}) $ in (\ref{ppo}) is the one based on the Laughlin charges of $1/2$ unit of charge, quite close and similar to the usual GMP mode at half-filling. 
 
\section{Heisenberg equations of motion of the pairing operator}

We model collective modes of the Pfaffian systems (with bosons or electrons) through the superpositions of Cooper pairs with non-zero center of mass momentum i.e. via the pairing operator in (\ref{ppo}). To derive Heisenberg equation of motion of this  operator we use the Hamiltonian in (\ref{modHam}), but in the form of the BCS reduction i.e. with a projection to the Cooper channel of the second term. Any single particle piece from the second part, together with any HF contribution that can be derived from the Hamiltonian is descibed by the usual single particle term in the BCS description. Thus, we have,
\begin{equation}
H_{\rm BCS} = H_0 + H_{\rm BCS}^{int}, \label{bcs}
\end{equation}
where
\begin{equation}
 H_0  = \sum_{\bs k} \xi_{\bs k} c_{\bs{k}}^\dagger c_{\bs{k}}, 
\end{equation}
and
\begin{equation}
 H_{\rm BCS}^{int} = \sum_{{\bs k}, {\bs q}}    e({\bs k}, {\bs q})  c_{\bs{k} - \bs{q}}^\dagger c_{-\bs{k} + \bs{q}}^\dagger c_{-\bs{k}} c_{\bs{k}}. \label{CI}
\end{equation}
In the  BCS Hamiltonian,  $\xi_{\bs k} =  \epsilon_{\bs k}  - \mu $, $\epsilon_{\bs k} $ - single particle dispersion, $\mu$ -- chemical potential, and 
\begin{eqnarray}
 e({\bs k}, {\bs q}) &= & C  \exp\left( i \frac{\bs{k} \times \bs{q}}{2}\right)  \times \nonumber \\
&& (\exp\left(- i \frac{\bs{k} \times \bs{q}}{2}\right) - \exp\left(i \frac{\bs{k} \times \bs{q}}{2}\right)).
\end{eqnarray}
The mean-field procedure and method will introduce the field $\Delta_{\bs k} $, in the interaction term,
\begin{equation}
H_{\rm BCS}^{mf} = H_0 + ( \sum_{\bs k} \frac{\Delta_{\bs k}^*}{2} c_{-\bs{k}} c_{\bs{k}} + h.c.),
\end{equation}
where
\begin{equation}
\frac{\Delta_{\bs k}^*}{2} =  \sum_{\bs q}   e({\bs k}, {\bs q})  \langle BCS|   c_{ - \bs{q}}^\dagger c_{\bs{q}}^\dagger |BCS\rangle , \label{ex1}
\end{equation}
and  $|BCS\rangle$ is the ground state of the problem defined  by the quadratic mean-field Hamiltonian. We also have
\begin{equation}
\frac{\Delta_{- \bs q}}{2} =  \sum_{\bs k}  e({\bs k}, {\bs q})   \langle BCS|    c_{ - \bs{k}} c_{\bs{k}} |BCS\rangle .\label{ex2}
\end{equation}
To study Heisenberg equations of motion for $\eta ({\bs{q}}) $ with respect to  the interaction operator in $H_{\rm BCS} $ i.e.  $ H_{\rm BCS}^{int}$, we begin with the following commutator,
\begin{eqnarray}
&&[\eta ({\bs{q}}) , H_{\rm BCS}^{int}] =  \nonumber \\
&&\sum_{{\bs k}, {\bs q}_1}  - 4  f(\bs{k}+ \bs{q}_1 , \bs{q})   e({\bs k}, - {\bs q}_1)  c_{\bs{k} + \bs{q}_1}^\dagger c_{\bs{k}+ \bs{q}_1 + \bs{q}} c_{-\bs{k}} c_{\bs{k}}.
\end{eqnarray}
We may try to introduce mean-field  $\Delta_{\bs k}$ instead of the superposition of $ c_{-\bs{k}} c_{\bs{k}}$ ooperators, in order to linearize the equations for $\eta ({\bs{q}}) $, but there is no simple decoupling in the expression on the r.h.s. and we see that the charge and pair excitations are entangled in a non-trivial way.

Therefore we study,
\begin{equation}
[[\eta ({\bs{q}}) , H_{\rm BCS}^{int}] , H_{\rm BCS}^{int}] ,
\end{equation}
i.e. the second derivative with respect to time, also, due to the expectation that the bosonic mode would obey an equation with the second derivative with respect to time. 
We get,
\begin{eqnarray}
&&[[\eta ({\bs{q}}) , H_{\rm BCS}^{int}] , H_{\rm BCS}^{int}] = \nonumber \\
&& \sum_{{\bs k}, {\bs q}_1}   \sum_{{\bs k}_1, {\bs q}_2} - 8  f(\bs{k}+ \bs{q}_1 , \bs{q})   e({\bs k}, - {\bs q}_1)  e({\bs k}_1 ,- {\bs q}_2)  \times \nonumber \\
&&  \{ \delta(\bs{k}+ \bs{q}_1 + \bs{q}, \bs{k}_1 + \bs{q}_2 ) \; c_{\bs{k} + \bs{q}_1}^\dagger c_{-\bs{k}_1 - \bs{q}_2 }^\dagger c_{-\bs{k}} c_{\bs{k}} c_{-\bs{k}_1} c_{\bs{k}_1} \nonumber \\
&&  -  \delta(-\bs{k}, \bs{k}_1 + \bs{q}_2 ) \; c_{\bs{k} + \bs{q}_1}^\dagger  c_{\bs{k}+ \bs{q}_1 + \bs{q}}    c_{-\bs{k}_1 - \bs{q}_2 }^\dagger           c_{\bs{k}} c_{-\bs{k}_1} c_{\bs{k}_1} \nonumber \\
&&  +  \delta(\bs{k}, \bs{k}_1 + \bs{q}_2 ) \;  c_{\bs{k} + \bs{q}_1}^\dagger  c_{\bs{k}+ \bs{q}_1 + \bs{q}}   c_{- \bs{k}}   c_{-\bs{k}_1 - \bs{q}_2 }^\dagger           c_{-\bs{k}_1} c_{\bs{k}_1} \nonumber \\
&&  - \delta(\bs{k}+ \bs{q}_1, - \bs{k}_1 ) \; c_{\bs{k} + \bs{q}_1}^\dagger c_{-\bs{k}_1 - \bs{q}_2 }^\dagger c_{\bs{k}_1} c_{\bs{k}+ \bs{q}_1 + \bs{q}} c_{-\bs{k}} c_{\bs{k}}\}.  \nonumber \\
&& 
\end{eqnarray}
It is not hard, by using expressions  (\ref{ex1}) and  (\ref{ex2})  for $\Delta_{\bs k}$ and (\ref{ppo})  for pairing operator to prove that the first term is proportional to $ \eta ({\bs{q}})^\dagger \Delta_{\bs k} \Delta_{{\bs k} + {\bs q}} $, and due to phases that come from the product  $  \Delta_{\bs k} \Delta_{{\bs k} + {\bs q}} \sim  \Delta_{\bs k}^2 $, its contribution may be negligible. Similarly, in the second and third term, we can come up with $\Delta_{{\bs q}_1} $ and no charge or pairing field can suppress phase fluctuations  associated with $\Delta_{{\bs q}_1} $.

On the other hand, for the contribution from the last term we get,
\begin{eqnarray}
&& \sum_{{\bs k}}   \sum_{{\bs k}_1, {\bs q}_2}  8  f(- \bs{k}_1 , \bs{q})   e({\bs k}, {\bs k}_1)  e({\bs k}_1 ,- {\bs q}_2)  \times \nonumber \\
&& \;\;\;\;\;\;\;   c_{\bs{k} + \bs{q}_1}^\dagger c_{-\bs{k}_1 - \bs{q}_2 }^\dagger c_{\bs{k}_1} c_{-\bs{k}_1  + \bs{q}} c_{-\bs{k}} c_{\bs{k}}\}  \nonumber \\
&& \approx  \sum_{{\bs k}_1, {\bs q}_2}  8   f(- \bs{k}_1 , \bs{q}) \frac{\Delta_{- {\bs k}_1}}{2} e({\bs k}_1 ,- {\bs q}_2)  \times \nonumber \\
&&\;\;\;\;\;\;\;  c_{\bs{k} + \bs{q}_1}^\dagger c_{-\bs{k}_1 - \bs{q}_2 }^\dagger c_{\bs{k}_1} c_{-\bs{k}_1  + \bs{q}}  \nonumber \\
&& \approx  \sum_{{\bs k}_1}  8  f( \bs{k}_1 , \bs{q})  \frac{\Delta_{{\bs k}_1}}{2} \frac{\Delta^*_{{\bs k}_1}}{2}  c_{-\bs{k}_1  + \bs{q}}  c_{\bs{k}_1}. \nonumber \\
&&
\end{eqnarray}
In the weak-coupling regime we can approximate all summations by summations around Fermi sphere. If we assume that $\Delta_{\bs k}$ is nearly constant in that region, we can write
\begin{equation}
[[\eta ({\bs{q}}) , H_{\rm BCS}^{int}] , H_{\rm BCS}^{int}]  \approx 2 |\Delta_{k_F} |^2  \eta ({\bs{q}}).
\end{equation}
In a similar fashion , using $\xi_{k_F} = 0$, the long distance limit, $|{\bs q}| \rightarrow 0$, and neglecting the contributions  due to oscillating phase, we reach the conclusion for the commutator with the complete Hamiltonian in (\ref{bcs}),
\begin{equation}
[[\eta ({\bs{q}}) , H_{\rm BCS}] , H_{\rm BCS}]  \approx 2 |\Delta_{k_F} |^2  \eta ({\bs{q}}).
\end{equation}
Thus, there are no two distinct modes, $\eta ({\bs{q}}) \pm \eta^\dagger  ({\bs{q}})$, with different dispersions, instead, we have the same equation for $\eta ({\bs{q}})$ and $\eta^\dagger  ({\bs{q}}) $, and thus a single bosonic mode. Therefore, in the case of the $p$-wave topological superconductor, in the weak-coupling regime, instead of two modes,  Goldstone and Higgs, that are present in the ordinary SC, we can expect only one mode.  This is in an agreement with the claim of Ref. \cite{jbcs} that two modes are uncoupled in the ordinary BCS SC due to the presence of a charge conjugation symmetry. That symmetry is absent in our case.

Moreover, we can claim the exact degeneracy and a single collective mode, based on the general expectation for the effective, Ginzburg-Landau description of the $p$-wave superconductor, we introduced via microscopic description in (\ref{CI}). Namely, in the case of the superconductor that explicitly breaks the time reversal symmetry, in the long-wavelength, low-energy description we expect the basic time-dependence to enter via the first derivative in time, just as in the usual case of ordinary superfluid, and thus describe the density and phase of the order parameter as conjugate variables, which entangled oscillations produce a single collective mode. Based on the microscopic calculations we recovered the dispersion (energy) associated with this single mode.

In a similar manner we can approach the physics of magnetophonon mode - the GMP mode in the CF representation. In the long-distance limit, we know that it represents a quadrupolar excitation with two pairs of Laughlin quasiparticle - quasihole excitations. Thus in both systems, either with electrons or bosons, we can associate this mode with the (physical) density operator in (\ref{BDen}), expressed via CFs that are made of these pairs. As in the case of $\eta ({\bs{q}})$ operator, after the same kind of approximations, we get,
\begin{equation}
[[\rho ^L({\bs{q}}) , H_{\rm BCS}] , H_{\rm BCS}]  \approx 4 |\Delta_{k_F} |^2  \rho^L ({\bs{q}}).
\end{equation}

\section{Single collective mode description and a evolution into a PDW}

The final forms of the Heisenberg equations in the previous section describe the energy quanta: the energy is $ \sqrt{2} |\Delta_{k_F} |$ for the pairing mode, and  $2  |\Delta_{k_F} |$ for the charge (GMP) mode. Thus we can concentrate on the subspace of the lowest lying collective mode in the long-distance, based on the SC  phase and pairing of CFs, connected with mixed, entangled amplitude and phase oscillations, similar to the phonon mode of superfluid. 

In the mean-field approximation we can assume that for each $ {\bs q} \neq 0$ we have a simple bosonic algebra,
\begin{equation}
[\eta ({\bs{q}}) , \eta ({\bs{q}})^\dagger ]  \approx 2 \sum_{{\bs p}}  \sin^2 \left(\frac{\bs{q} \times \bs{p}}{2}\right)  (1 - \langle c_{\bs p}^\dagger c_{\bs p}\rangle ) = F(q).
\end{equation}
In the following we will redefine $\eta ({\bs{q}}) $, as  $\eta ({\bs{q}})/\sqrt{F(q)} \rightarrow b_{\bs q} $, in order
 to work with operators,  $b_{\bs q} $ and  $b_{\bs q}^\dagger $ , that take the roles of bosonic annihilation and creation operators. The normalized states of the subspace defined by the collective excitaions of the SC are
\begin{equation}
b_{{\bs q}_1}^\dagger  \cdots  b_{{\bs q}_n}^\dagger  |BCS \rangle  \label{subs}.
\end{equation}
we can express tha Hamiltonian in that subspace as
\begin{equation}
H = \sum_{\bs q} \delta E  b_{{\bs q}}^\dagger b_{{\bs q}}  \label{Hsubs},
\end{equation}
where $ \delta E  = \sqrt{2} |\Delta_{k_F} |$ is the energy of the excited mode. Because any  state as in (\ref{subs}) represents a superposition of an even (doubled) number of CFs, we may write
\begin{equation}
 \sum_{\bs q}   b_{{\bs q}}^\dagger b_{{\bs q}}  = 1/2   \sum_{\bs k}   c_{{\bs k}}^\dagger c_{{\bs k}}  \label{num},
\end{equation}
Thus, in this subspace,
\begin{equation}
H = \sum_{\bs k}\frac{ \delta E}{2}  c_{{\bs k}}^\dagger c_{{\bs k}},
\end{equation}
 Also, we must be aware that we work in the long-distance limit.  If we expose the system to conditions that will induce ordering in the system, in the real space, we expect rearrangement of states in the long-distance sector that will lead to a gapless Goldstone mode.

If we assume that the states of the collective mode, due to the  $U(1)$ symmetry breaking, assciated with pairing states, are decoupled from the rest of the system dynamics in the long-distance limit  (and thus possibly weakly-coupled in reality and close to the notion of quantum scars) we find that the only perturbation that will act in this space and still maintain the quadratic Hamiltonian is the one that describes enetering and leaving of a quadrupole. (Thus the only redistribution of the matter in the long-distance will be quadrupolar excitations, as a relevant perturbation.) We choose a definite, non-zero momentum of the center of mass of the quadrupolar excitation, and thus a specific inhomogeneous ordering in the real space. Thus our Hamiltonian is
\begin{equation}
H = \sum_{\bs q}\frac{ \delta E}{2}  c_{{\bs q}}^\dagger c_{{\bs q}} + \gamma  \eta^\dagger ({\bs{Q}}) + \gamma^* \eta ({\bs{Q}}). \label{etaH}
\end{equation}
By adopting this form of the Hamiltonian we are also assuming that the vector ${\bs{Q}}$ is not too large to shift momenta from the region around $ {\bs{q}} = 0$ i.e. long-distance domain where the dispersion can be approximated by the constant,  $(\delta E)/2 $. 

The description and definition of a PDW, in a system of electrons in a superconducting phase, entails the presence of both, ${\bs{Q}}$  and $ - {\bs{Q}}$  vectors \cite{pdwreview}, with no net supercurrent. But in the present case of FQHE we may allow for a single vector ${\bs{Q}}$, because the superconducting behaviour is on the level of the composite quasiparticles - CFs, which, in the first approximation are neutral particles and, furthermore, in a LL, no real current can exist. In a mean-field picture of the ensuing ground state, in the first approximation, we can neglect the backflow corrections, and consider the state with definite ${\bs{Q}}$. On the other hand, that does not preclude the possibility that in order to preserve the symmetry under inversion \cite{halI}, the system has both, ${\bs{Q}}$  and $ - {\bs{Q}}$. In a mean-field, BCS picture, we would have two decoupled condensates (of Cooper pairs of CFs), with opposite momenta. The problem defined in (\ref{etaH}) will be essentially doubled, and thus we may just concentrate on the solution of (\ref{etaH}).

The Hamiltonian in (\ref{etaH})  is very similar to the one discussed in \cite{pok} in the description of the quantum scars in the context of lattice models with spin. As in that case, here we have a (spinless) system in a strong-pairing phase. In our case $ \mu < 0$, in the notation of  \cite{rg}, and after going through the same algebra we can conclude that for the ground state we have, in the long-distance limit, the following expression,
\begin{equation}
| \Psi_0\rangle \sim  \exp(- \frac{\gamma}{\delta E} \; \eta^\dagger ({\bs{Q}}) ) |0\rangle .
\end{equation}
Thus we have a ground state where each Cooper pair has the non-zero momentum  ${\bs{Q}}$, of the center of mass of the pair. In a PDW state, both ${\bs{Q}}$ and $ - {\bs{Q}}$ are present,
\begin{equation}
| \Psi_0\rangle^{PDW}  \sim  \exp(- \frac{2 \gamma}{\delta E} \; \eta^\dagger ({\bs{Q}}) ) |0\rangle  \exp(- \frac{2 \gamma}{\delta E} \; \eta^\dagger (- {\bs{Q}}) ) |0\rangle .
\end{equation}

\section{Discussion and Conclusions}
In short, we proposed an alternative scenario for  a PDW state for FQHE at 5/2 (even-denominator filling) as a possible precursor state and explanation for the nematic behaviour seen in experiments \cite{csathy}, with increasing pressure, in the two-dimensional system. The weak-coupling regime and evolution of the system (in the framework of our microscopic approach), we believe is the most natural setting and description of the induced change in the system as the pressure is varied. In this way, ensuing a strong-pairing scenario for a PDW state may be  more likely.  

Furthermore, the proposed scenario may be relevant for other systems, layered or twisted new materials, that are expected to support the chiral ($p$-wave) superconducting phase. The chiral interaction, as encoded in the Hamiltonian we worked with (Eq. (\ref{CI})), in its long-distance, low-energy form,  is likely a universal description of paired electrons in new materials. The presence of a single collective mode in these systems (in the long-distance physics) in an assumed weak-coupling regime of the superconducting state, facilitate the transition into a PDW state, and in that sense the phase may be prone to such, additional ordering.

On the other hand, strong-coupling, weak-pairing scenario of Ref. \cite{frad} may be directly related to critical points and behaviour that can be resolved (in the neighborhood of critical points or regions)  by stabilizing  into a strong-coupling, weak-pairing $p$-wave state of CFs  (Pfaffian states) next to non-uniform states. One example  is the transition into the nematic state of the system of bosons at $\nu = 1$, by increasing the  Haldane $V_2$ pseudopotential with respect to $V_0$ \cite{b3}. The other example is the system with electrons of the half-filled second Landau level, when the Haldane $V_1$ pseudopotential is decreased from the Coulomb value \cite{rh}. In both cases the Pfaffian ordering is the strongest near the transition, and its existence may be connected with critical behaviour \cite{nnmvm}.
 \section{Acknowledgments}  We thank Ch. Varma for discussions.   We acknowledge funding provided by the Institute of Physics Belgrade, through the grant by the Ministry of Science, Technological Development, and Innovations of the Republic of Serbia.
This work was performed in part at the Aspen Center for Physics, which is supported by National Science Foundation grant PHY-2210452.
\appendix*
\section{Derivation of pairing operator}

In this appendix we will derive the pairing operator, $\eta ({\bs{q}}) $ (Eq.(\ref{ppo})) starting from the basic set-up and formalism developed for CFs in Refs. \cite{read, ph}, in the case of bosons at filling factor $\nu = 1$. The same formalism can be extended  and applied to the system  of half-filled Landau level of electrons \cite{ms, nnmvm, pkm}. 

The continuum, field-theoretical description \cite{ph,read} starts with a two-index fermionic field, $c_{m n}$, in an enlarged space, where indexes, $ m, n = 1, ... N_\phi  $ denote the states of a chosen basis in a LL.  $ N_\phi $ denotes the number of flux quanta through the system. The left index, $m$, denotes the state of boson and the right index, $n$, denotes the state of its correlation hole,  of the composite object - composite fermion described by field $c_{m n}$, $c_{m n}^\dagger $;
\begin{equation}
\{c_{m n}, c_{n' m'}^{\dagger}\}= \delta_{n,n'} \delta_{m,m'}.
\end{equation}
The density operators that may be introduced are: ''physical'' density,
 \begin{equation}
\rho_{n n'}^{L} = \sum_{m} c_{m n}^{\dagger} c_{n' m},
\label{LDen}
\end{equation}
i.e. the density of bosons, and artificial, ''unphysical'' density,
\begin{equation}
\rho_{m m'}^{R} = \sum_{n} c_{m n}^{\dagger} c_{n m'},
\label{RDen}
\end{equation}
refering to the density of correlation holes. 

Given that there are as many correlation holes as bosons, and that correlation holes have fermionic statistics, the necessary constraint that will reduce the number of the degreees of freedom (of the enlarged space) to the physical ones is
\begin{equation}
 \rho_{n n}^R = 1 . \label{nncon}
\end{equation}
On the other hand, one may introduce the description in the inverse space, 
\begin{equation}
c_{\bs{k}}= (2 \pi)^{\frac{1}{2}} \sum_{m,n}  \langle n\vert\tau_{-\bs{k}}\vert m\rangle   c_{mn},
\end{equation}
with $\tau_{\bs{k}} = \exp\left(i \bs{k} \cdot\bs{R}\right)$, where $\bs{R}$ is a guiding-center coordinate of a single
particle, of charge $q = - e < 0$, in the external magnetic field, $ \bs{B} = - B {\bs{e}_z} $, so that
\begin{equation}
[R_x , R_y ] = - i ,
\end{equation}
where we took $l_B$ (magnetic length) $=1$. Thus the composite object - dipole will have momentum ${\bs k}$, if the two (localized) states of the composite are distance $ {\bs e_z} \times {\bs k} $ apart, because $ {\bs e_z} \times {\bs R} $ has the role of translation operator  in a LL.  The inverse relationship exists and incudes an integration over the complete $ {\bs k} -$plane:
\begin{equation}
c_{nm} = \int\frac{d\bs{k}}{(2\pi)^{\frac{3}{2}}}        \langle n\vert\tau_{\bs{k}}\vert m\rangle        \;           c_{\bs{k}}.  \label{excnm}
\end{equation}
The same expression can be used in the formalism for densities in (\ref{LDen}) and (\ref{RDen}), to recover the expressions for operators in the inverse space in (\ref{BDen}) and (\ref{BDenr}).

We can introduce the pairing operator in an analogous way,
 \begin{equation}
\eta_{m m'} = \sum_{n} c_{m n}  c_{n m'}.
\end{equation}
By substituing  (\ref{excnm}) we get in the inverse space,
\begin{eqnarray}
&&\eta_{m m'}  =  \nonumber \\
&=& \int\frac{d\bs{k}_1}{(2\pi)^{\frac{3}{2}}}  \int\frac{d\bs{k}_2}{(2\pi)^{\frac{3}{2}}}    \langle m\vert\tau_{\bs{k}_1}\tau_{\bs{k}_2}  \vert m'\rangle    c_{\bs{k}_1}  c_{\bs{k}_2} \nonumber \\
&=& \int\frac{d\bs{q}}{(2\pi)^{\frac{3}{2}}}  \int\frac{d\bs{k}}{(2\pi)^{\frac{3}{2}}}    \langle m\vert\tau_{\bs{q}}  \vert m'\rangle   \exp\left( i \frac{\bs{q} \times \bs{k}}{2}\right)   c_{- \bs{k} +\bs{q} }  c_{\bs{k}} \nonumber \\
&=& \int\frac{d\bs{q}}{(2\pi)} \langle m\vert\tau_{\bs{q}}  \vert m'\rangle   \eta ({\bs{q}}), \nonumber \\
\end{eqnarray}
where
\begin{equation}
\eta ({\bs{q}}) = \int\frac{d\bs{k}}{(2\pi)^{2}}  \exp\left( i \frac{\bs{q} \times \bs{k}}{2}\right)  c_{- \bs{k} +\bs{q} }  c_{\bs{k}}.
\end{equation}
Using an invariance under the following change of variable, $ \bs{k} \rightarrow \bs{k} + \bs{q}$ we get the final form in the main text. 

In essence, we take the point of view of Murthy and Shankar \cite{ms} that the additional degrees of freedom are collective, physical degrees of freedom, similar to the work of Bohm and Pines \cite{bp} with plasmons in the context of electron gas, and we treat them as well defined entities in the theory.  As in the case of plasmons and Coulomb gas, the "correlation holes" are physical entities connected with higher energies, and as such take part in the building of correlations in the ground state. But considering and introducing them as well-defined degrees of freedom, we have to be careful and ensure that the number of degrees of freedom stays the same and thus constraints. Therefore, these additional holes, vortices are not external, auxiliary variables, they are a part of the physical space, and the excitonic correlations and attraction is a way to understand the pairing instability at the mean-field level. The way we introduced the pairing operator reflects this point of view. On the other hand this is the most natural if not the only way in the theory that we can introduce a particle-particle CF bilinear.


\end{document}